%% file: main.tex
\documentclass[conference,a4paper]{IEEEtran}
\usepackage[left=1.43cm,right=1.43cm,top=1.82cm,bottom=4.3cm]{geometry}

\usepackage[algo2e]{algorithm2e} 
\usepackage{amsmath,amssymb,amsmath,amsfonts}
\usepackage{bm}
\usepackage{bbm}
\usepackage{graphicx}
\usepackage{cite}
\usepackage{epstopdf}
\usepackage{gensymb}
\usepackage{color}
\usepackage{lipsum}
\usepackage{float}
\usepackage{epstopdf}
\usepackage[mathcal]{eucal}
\usepackage{subfiles}
\usepackage[T1]{fontenc}
\usepackage{siunitx}
\usepackage{tabulary}
\usepackage{epsfig,graphics,graphicx,subfig,amssymb,amstext,amsmath,algorithm,algorithmic,wrapfig,multirow}
\usepackage{array}
\usepackage{adjustbox}
\usepackage[dvipsnames]{xcolor}
\usepackage{cite}
\usepackage{times}
\usepackage{placeins}
\usepackage{textcomp}
\usepackage[font=small]{caption}
\usepackage{flushend}
\usepackage{color, colortbl}
\usepackage{tabularx}
\usepackage{bm}
\usepackage{enumerate}
\usepackage{url}
\usepackage{tikz}
\usepackage{pgfplots}
\usepackage{epstopdf}
\usetikzlibrary{shapes,arrows}
\usepackage[utf8]{inputenc}
\usepackage{mathtools}
\usepackage[utf8]{inputenc}
\usepackage{xcolor}
\usepackage{tikz}
\usetikzlibrary{chains,arrows,calc,positioning}
\usepackage{amssymb}
\usepackage{pifont}
\usepackage{soul}
\usepackage{breqn} 
\usepackage{booktabs} 
\usepackage{multirow}
\usepackage{enumitem}
\usepackage{tabulary}
\usepackage[normalem]{ulem}
\usepackage{dblfloatfix}
\usepackage{makecell}
\usepackage{lipsum}
\usepackage{amsthm}
\usepackage{comment}
\usepackage{filecontents}

\newtheoremstyle{mystyle}
  {}
  {}
  {\itshape}
  {}
  {\bfseries}
  {.}
  { }
  {}
\theoremstyle{mystyle}

\newlength \figwidth
\definecolor{bittersweet}{rgb}{1.0, 0.44, 0.37}
\definecolor{glaucous}{rgb}{0.38, 0.51, 0.71}
\definecolor{gainsboro}{rgb}{0.86, 0.86, 0.86}
\definecolor{babyblueeyes}{rgb}{0.63, 0.79, 0.95}
\definecolor{silver}{rgb}{0.75, 0.75, 0.75}
\definecolor{neoncarrot}{rgb}{1.0, 0.64, 0.26}
\definecolor{Gray}{gray}{0.9}
\definecolor{LightCyan}{rgb}{0.88,1,1}
\definecolor{BackgroundLightBlue}{rgb}{0.97,0.97,1}
\definecolor{BackgroundGray}{gray}{0.98}

\makeatletter

 \let\oldforeign@language\foreign@language
 \DeclareRobustCommand{\foreign@language}[1]{%
   \lowercase{\oldforeign@language{#1}}}

\input{commands.tex}

\makeatother
\IEEEoverridecommandlockouts
\begin{document}

\bstctlcite{IEEEexample:BSTcontrol}

\title{Site-Specific RIS Deployment in Cellular Networks via Calibrated Ray Tracing}

\author{\IEEEauthorblockN{Sina Beyraghi$^{\dagger\,\star}$, Javad Shabanpour$^{\flat}$, Giovanni Geraci$^{\star}$, Paul Almasan$^{\dagger}$, and Angel Lozano$^{\star}$ \vspace{0.1cm}
}
\\ \vspace{-0.3cm}
\normalsize\IEEEauthorblockA{$^{\star}$\emph{Univ. Pompeu Fabra, Barcelona, Spain} \enspace \enspace $^{\dagger}$\emph{Telefónica Research, Barcelona, Spain}  \\   $^{\flat}$\emph{Nokia Bell Labs, Espoo, Finland}}


\thanks{
This work was in part supported by H2020-MSCA-ITN-2020 META WIRELESS (Grant Agreement: 956256), by the SNS
JU Horizon Europe Project under Grant Agreement No.
101139161 (INSTINCT), by the 6G-Machine Intelligence based Radio Access Infrastructure (6G-MIRAI) project under Grant Agreement No. 101192369, by the Spanish Research Agency through grants PID2021-123999OB-I00 and CNS2023-145384, by ICREA, by the Maria de Maeztu Units of Excellence Programme (CEX2021-001195-M), and by the Spanish Ministry of Economic Affairs and Digital Transformation and the European Union NextGenerationEU through UNICO-5G I+D projects TSI-063000-2021-138 (SORUS-RIS) and TSI-063000-2021-59 (RISC-6G).}

} 

\maketitle

\input{00_Abstract}
\input{01_Introduction}
\input{02_SystemModel}

\input{03_Calibration}
\input{04_Algorithms}

\input{05_Results}

\input{06_Conclusion}
\bibliographystyle{IEEEtran}
\bibliography{journalAbbreviations, main}

\end{document}

%% file: commands.tex
\def\nb0{{\mathbf{0}}}
\def\nb1{{\mathbf{1}}}



%% file: 00_Abstract.tex
\begin{abstract}
This work introduces a fully-automated RIS deployment strategy validated through a digital twin, powered by Sionna ray tracing, of a UK city.
On a scene calibrated with measured data, the method jointly optimizes RIS placement, orientation, configuration, and BS beamforming across 4G, 5G, and hypothetical 6G frequencies. Candidate RIS sites are identified via scattering-based rays, while user clustering reduces deployment overhead. Results show that meaningful coverage enhancement requires dense, large-aperture RIS deployments, raising questions about the practicality and cost of large-scale RIS adoption.
\end{abstract}

%% file: 01_Introduction.tex
\section{Introduction}
\label{sec:Sec1}

Consistent connectivity is a key challenge in modern wireless systems, especially in dense urban environments where obstructions and multipath propagation degrade performance~\cite{10574348}. These issues worsen at higher frequencies due to increased pathloss and limited diffraction. Reconfigurable intelligent surfaces (RIS) have emerged as a promising solution. By reflecting signals toward underserved areas without additional transmit power, RIS offer a potentially energy-efficient means to enhance coverage~\cite{9140329,8741198}. However, large-scale adoption by mobile network operators (MNOs) remains uncertain due to unclear cost-performance tradeoffs. 

Indeed, RIS placement involves a high-dimensional combinatorial problem, with many candidate sites, orientation constraints, and site-specific propagation. Trial-and-error methods are unscalable.
Tackling this challenge requires an automated and scalable framework, and an enticing enabler of such a framework is a digital twin providing 
a precise digital replica of the radio environment \cite{10742568}.

Previous studies on RIS placement and design offer valuable insights but suffer from key limitations. Many focus on single-user equipment (UE) or single-cell scenarios using idealized or stochastic models that are unable to capture site-specific details~\cite{9530750, 9712623, 9351782, 9852464, 10211255}. Others use simplified reflection models that ignore polarization and angular losses~\cite{ 9140329, 10473672, sandh2024ray, 10301521}, or neglect practical constraints like building geometry, material properties, and orientation feasibility~\cite{9745477, 9586067}. Few combine ray tracing with empirical data, and none offer a scalable, automated solution across multiple frequencies and network layouts. 

This paper presents a fully automated, data-driven framework for site-specific RIS deployment in cellular networks. A radio digital twin is applied to jointly optimize RIS placement, orientation, phase configuration, and base station (BS) beamforming. Unlike analytical or heuristic methods, the framework supports multi-cell, multi-band deployments and incorporates a calibrated ray-tracing engine to model material- and geometry-specific electromagnetic interactions.

The framework is validated using data from a UK commercial deployment. Simulations span 4G, 5G, and a hypothetical 6G system at $2$\,GHz, $3.5$\,GHz, and $10$\,GHz, respectively, using the open-source Sionna engine~\cite{hoydis2023sionna}. Material properties are calibrated with measured UE received power.
The main contributions are as follows:

\begin{itemize}
\item Using measured UE data, material properties are calibrated to improve ray-tracing accuracy. 
\item A ray-based method jointly optimizes RIS placement, phase configuration, and BS beamforming, incorporating orientation and geometry constraints.
\item Coverage improvements are evaluated across frequencies as a function of RIS aperture, element count, and density.
\item A cost-performance analysis reveals that significant coverage gains require extensive deployment of large RIS units, questioning economic viability.
\end{itemize}

To foster reproducibility, the full simulation framework and RIS deployment algorithms are available in open source.\footnote{\url{https://github.com/Telefonica-Scientific-Research/DDRD}}

%% file: 02_SystemModel.tex
\section{Network, RIS, and Propagation Models}
\label{sec:Sys_Model}

\subsection{Cellular Network Model}

\begin{figure}[t]
    \centering
    \includegraphics[width=0.95\columnwidth]{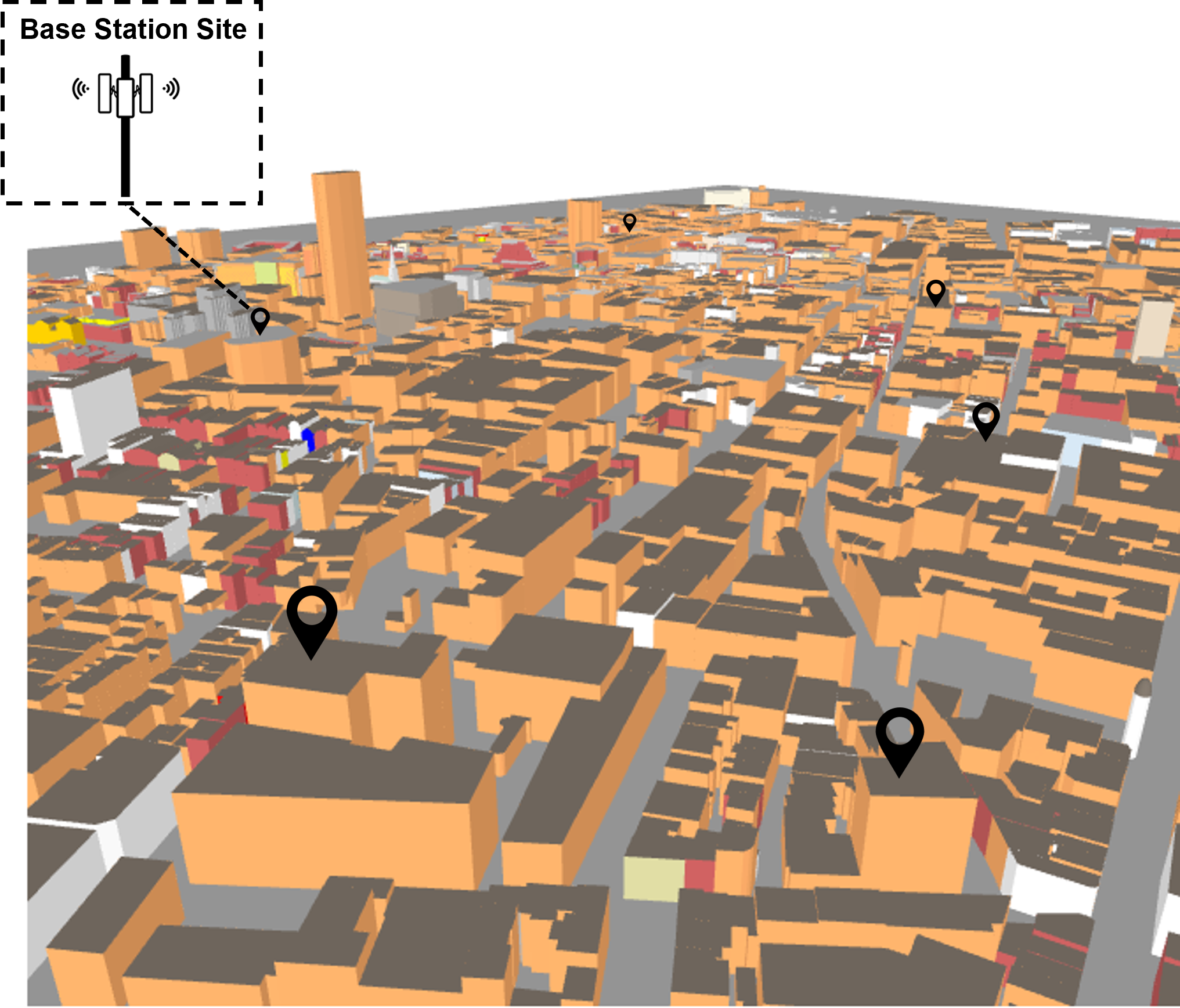}
    \caption{3D visualization of the area produced with OpenStreetMap, with black pins indicating BS sites.}
    \label{fig:3D_visualization}
\end{figure}

The digital twin replicates the layout of a commercial cellular network deployed by a leading MNO. The focus area contains 12 BSs ranging from 18 to 56\,m in height, each hosting three sectors for a total of 36 cells. The area spans 1340\,m~$\times$~1390\,m in the UK. Figure~\ref{fig:3D_visualization} shows a 3D rendering of the environment. 

\subsubsection{Antennas} 

Each BS features a planar array of vertically polarized elements. 
The configuration of each BS is specified by its vertical tilt angle and its horizontal bearing angle, both determined by the MNO.
A single radio-frequency chain feeds each antenna element and it abides by the 3GPP radiation pattern, with half-power beamwidths of $65^{\circ}$ in azimuth and $10^{\circ}$ in elevation~\cite{3GPP38901}. 

\subsubsection{Radio Deployments} 

Three systems are considered:
\begin{itemize}
    \item 4G at 2\,GHz (FR1)
    \item 5G at 3.5\,GHz (FR1)
    \item 6G at 10\,GHz (FR3)
\end{itemize}

\noindent For each, the parameters (adopted from commercial products, either existing or under development) are detailed in Table~\ref{tab:network_characteristics}. Note that only single polarization is assumed, resulting in element counts that are half those for dual polarization.
Reusing existing site grids is essential for the incorporation of new spectrum to be economically feasible, as further sites would lead to considerable expenses and prolong deployment timelines~\cite{holma2021extreme}. Hence, The same BS coordinates are applied to all three systems.\footnote{The 6G system operates at a higher frequency and with more bandwidth than its 4G/5G counterparts, but it is configured with a lower transmit power. This reflects a role for 6G as a non-standalone capacity-boosting layer with possibly discontinuous coverage.}

\begin{table}[ht]
    \centering
    \caption{System features \cite{lopez2024capacity} 
    }
    \footnotesize 
    \renewcommand{\arraystretch}{1.2} 
    \resizebox{0.9\columnwidth}{!}{ 
    \begin{tabular}{|l|c|c|c|}
        \hline
        \textbf{Feature} & \textbf{4G} & \textbf{5G} & \textbf{6G} \\
        \hline
        Carrier frequency [GHz] & 2 & 3.5 & 10 \\
        \hline
        Bandwidth [MHz] & 20 & 100 & 200 \\
        \hline
        Sectors per site & 3 & 3 & 3 \\
        \hline
        Polarization & Vertical & Vertical & Vertical \\
        \hline
        Planar array topology & 2×2 & 4×8 & 4×16\\
        \hline
        Beamforming codebook size & 4 & 32 & 64 \\
        \hline
        TX power per cell [dBm] & 43 & 49 & 44 \\
        \hline
        TX power per subcarrier [dBm] & 12.2 & 13.85 & 8.85 \\
        \hline
        Number of subcarriers & 1200 & 3276 & 3276\\
        \hline
        Noise power per subcarrier [dBm] & -132  & -129  & -126 \\
        \hline
    \end{tabular}
    } 
    \label{tab:network_characteristics}
\end{table}

\subsubsection{Beamforming Codebook} 

Each BS adopts a two-dimensional discrete Fourier transform (DFT) beamforming codebook per system. This codebook specifies a set of orthogonal beams on given angular directions; each such beam is characterized by the Kronecker product of two one-dimensional DFT vectors, one for azimuth and one for elevation, namely
\begin{equation}
    \boldsymbol{w}_{m_{\text{h}}, m_{\text{v}}} = \boldsymbol{w}_{m_{\text{h}}} \otimes \boldsymbol{w}_{m_{\text{v}}},
    \label{eq:DFT_UPA}
\end{equation}
where \( m_{\text{h}}, m_{\text{v}} \) are the horizontal and vertical beam indices. 

\subsubsection{UEs}

UEs are deployed on a grid of 2\,m~$\times$~2\,m tiles at a height of 1.5\,m, with the received power averaged per tile. The focus is on outdoor UEs as the main use case for cellular networks, though the method extends to indoor scenarios given floor plans and outdoor-to-indoor ray tracing support. Each UE employs a single vertically polarized isotropic antenna and is served individually per time-frequency resource, isolating the effects of scheduling.

\subsection{ Physically consistent RIS Model}
\label{sec:RISmodel}

The physically consistent RIS model in~\cite{9713744} is adopted, which treats the RIS as a continuous reradiating surface with a spatial modulation function based on ray optics. This captures finite surface size, angular reradiation, and energy conservation. Precisely, the spatial modulation function \( \Gamma(x_{\text{R}}, y_{\text{R}}) \) describes the complex transformation applied by the RIS at each surface point $(x_{\text{R}},y_{\text{R}})$, combining amplitude and phase modulation into \cite{10419169}

\begin{equation}
    \Gamma(x_{\text{R}},y_{\text{R}}) = R \sqrt{\eta} A(x_{\text{R}},y_{\text{R}}) e^{j \varphi(x_{\text{R}},y_{\text{R}})},
\end{equation}
where \( R \) accounts for roughness and loss, \( \eta \) is the surface efficiency, and \( A(x_{\text{R}},y_{\text{R}}) \) controls reradiated power. 

\subsection{Site-specific Ray Tracing and Coverage Calculation}

The propagation among BSs, UEs, and RIS units is digitally replicated using Sionna~\cite{10705152}, an open-source ray-tracing simulator. To balance realism and efficiency, the ray tracing employs a Fibonacci shoot-and-bounce method with $10^7$ rays per cell, up to four bounces per ray, and it includes specular reflection, diffraction, and scattering. 

To compute the large-scale channel gain between each BS and UE, the set of corresponding rays is considered. 
As advanced, the region of interest is partitioned into square tiles \( C_{p,q} \) of size 2\,m $\times$ 2\,m, where each tile corresponds to a UE. For each BS sector \( t \) and DFT beam index \( m  \), the directional channel gain at tile \((p,q)\) is computed as
\begin{equation}
    G_{t,m,p,q} = \frac{1}{4}  \iint_{C_{p,q}} \left| \boldsymbol{h}_{t,p,q}^*(x,y) \boldsymbol{w}_{t,m} \right|^2 \, dx dy,
    \label{eq:path_gain}
\end{equation}
where \( \boldsymbol{h}_{t,p,q}(x,y) \) is the channel vector at position \( (x,y) \), obtained from all ray types (LoS, reflected, diffracted, and scattered), \( \boldsymbol{w}_{t,m} \) is the DFT precoding vector, and \( 4\) is the tile area.

The reference signal received power (RSRP) for tile \( (p,q) \), served by BS sector \( t \) and beam index \( m \), is then 
\begin{equation}
    \text{RSRP}_{p,q} = G_{t,m,p,q} P_{{\text{t}}},
    \label{RSS_computation}
\end{equation}
where \( P_{\text{t}} \) is the transmit power per subcarrier, as specified in Table~\ref{tab:network_characteristics}.
The UE at the center of each tile selects its serving BS and beam by maximizing the RSRP across all sectors and beam indices.

%% file: 03_Calibration.tex
\section{Scene Material Calibration}
\label{sec:Calib}

Accurate ray tracing requires not only precise geometry but also realistic electromagnetic material properties—something particularly difficult at city scale. While prior work focused on small, controlled environments \cite{10705152}, we target dense urban areas where default material settings lead to significant mismatches. Using a large outdoor measurement dataset from a UK city---covering BS configurations (e.g., orientation, tilt) and spatial received power---material parameters are calibrated to align simulated and experimental radio coverage.

Specifically, building material parameters are calkibrated by minimizing the difference between simulated and measured RSRP values. Using the Sionna ray tracer and the actual BS configurations, coverage is first simulated assuming all buildings are made of concrete. Three material properties---relative permittivity, conductivity, and surface scattering---are treated as learnable variables. UE measurement data is aggregated over \(10 \times 10\,\mathrm{m}^2\) regions with at least 20 samples, enabling stable RSRP comparisons. For each region, surrounding buildings within 100 meters are jointly calibrated by comparing average simulated and measured RSRP. Gradients are computed using an Adam optimizer, and materials are iteratively updated. The optimization runs for 600 steps per cell, gradually covering the city while preserving already calibrated areas. To ensure meaningful updates, outliers and regions with persistent mismatches are filtered out. Region merging and error thresholds are applied to manage complexity and adapt to urban density variations.

The design of this calibration approach is shaped by both modeling simplifications and practical data limitations. Since the focus is on minimizing RSRP error rather than retrieving physically exact material values, the optimizer adjusts parameters to best match the model’s predictions with measurements. Calibration is performed over spatial regions instead of individual samples to mitigate device-related RSRP variability. 
Regions with extreme mismatches are excluded, as these are typically due to missing scene geometry. The optimizer reacts adaptively: if simulated RSRP overshoots, material properties are driven toward free-space conditions; if it undershoots, reflectivity is increased; in both cases, within constrained bounds. Persistent errors at the extremes trigger region exclusion. And, to manage urban complexity, adjacent buildings are merged, and region-specific thresholds are applied based on density, acknowledging that calibration tolerance varies across the cityscape.

\subsection{Validation}
\begin{figure}[h]
    \centering
    \includegraphics[width=0.9\columnwidth]{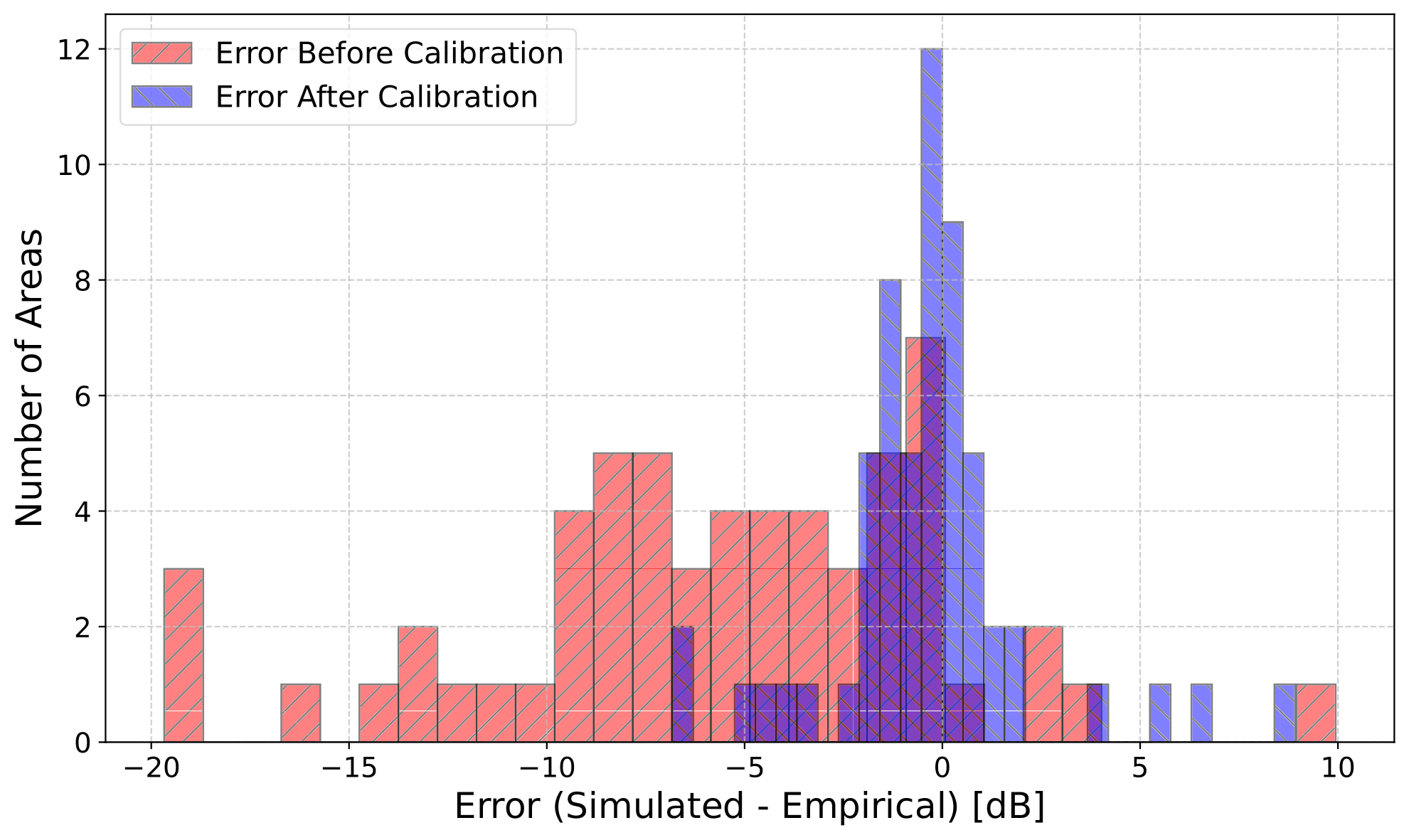}
    \caption{Distribution of RSRP prediction errors.}
    \label{fig:cal_histogram}
\end{figure}

Material calibration was conducted over a central urban area measuring 
\(1122\,\mathrm{m} \times 710\,\mathrm{m}\). To assess its effectiveness, 70 representative target regions were selected across the area. In each region, the average measured RSRP was compared against simulated values, both before and after calibration. A detailed analysis of prediction errors is provided in Fig.\ref{fig:cal_histogram}, with outlier regions excluded. Prior to calibration, RSRP predictions tended to underestimate coverage, with a mean error \(-5.69\,\mathrm{dB}\), median of \(-5.37\,\mathrm{dB}\), and standard deviation of \(5.71\,\mathrm{dB}\). After calibration, the error distribution narrowed substantially, with a near-zero mean of 
\(-0.32\,\mathrm{dB}\), median of \(-0.13\,\mathrm{dB}\), and standard deviation of \(2.57\,\mathrm{dB}\), confirming the effectiveness of the calibration approach.

%% file: 04_Algorithms.tex
\section{Data-Driven RIS Deployment}
\label{sec:Alg}

This section presents the data-driven strategy for large-scale RIS deployment on the digital twin.

\begin{figure*}[t]
    \centering
    \subfloat[4G deployment at 2\,GHz]{%
        \includegraphics[width=0.32\textwidth]{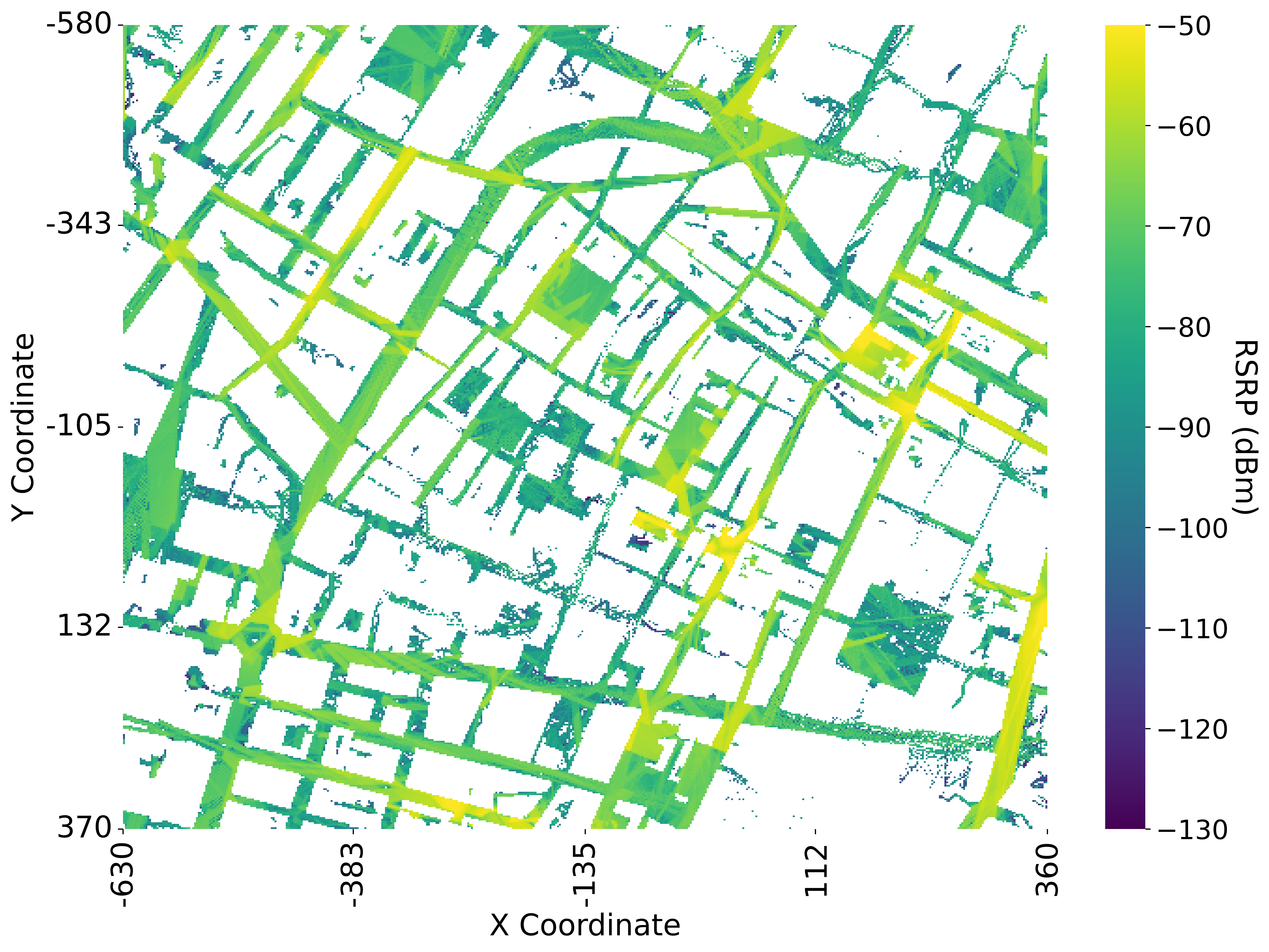}
        \label{fig:Heatmap_2GHz}
    }
    \hfill
    \subfloat[5G deployment at 3.5\,GHz]{%
        \includegraphics[width=0.32\textwidth]{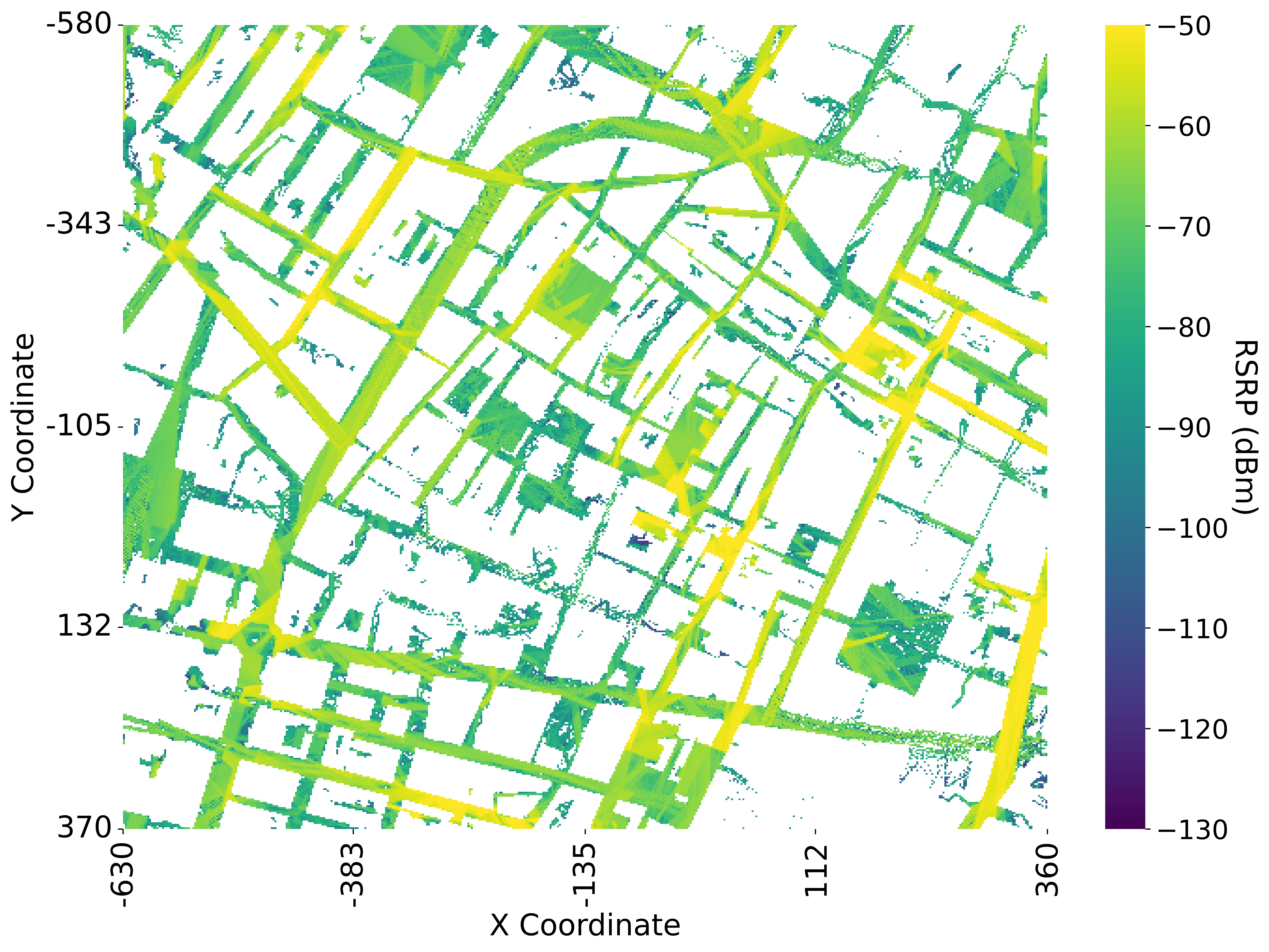}
        \label{fig:Heatmap_3_5GHz}
    }
    \hfill
    \subfloat[6G deployment at 10\,GHz]{%
        \includegraphics[width=0.32\textwidth]{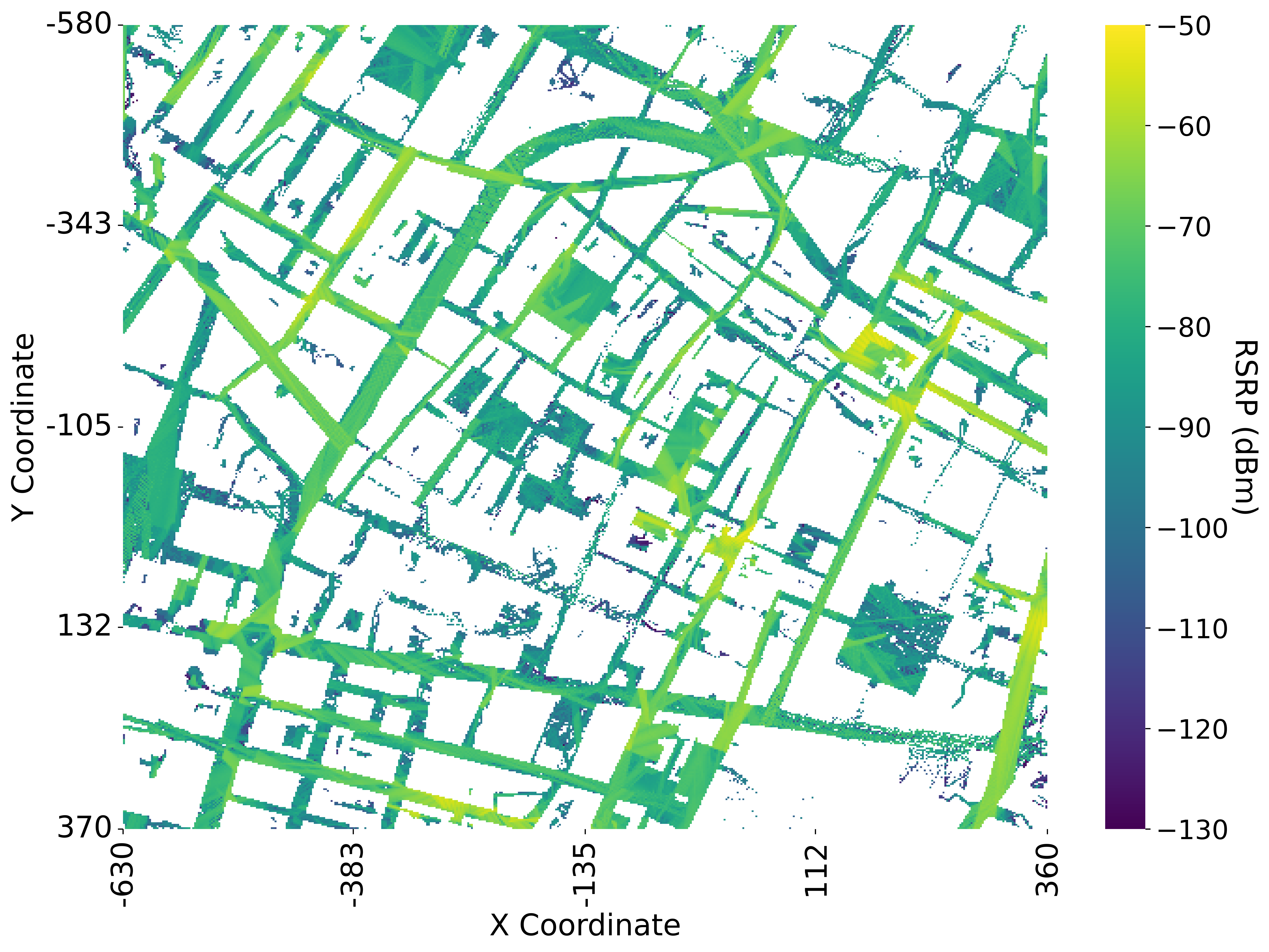}
        \label{fig:Heatmap_10GHz}
    }
    \caption{RSRP heatmaps across the considered urban area for 4G, 5G, and 6G.}
    \label{Heatmap}
\end{figure*}

\subsection{Coverage Evaluation and UE Clustering}

The baseline network coverage is evaluated by deploying BSs and distributing UEs on a grid, as described in Sec.~\ref{sec:Sys_Model}. Fig.~\ref{Heatmap} shows the RSRP heatmaps. UEs with an RSRP below $-100$\,dBm are declared in outage~\cite{teltonika}, resulting in outage rates of 2.85\%, 1.65\%, and 6.07\% at 2\,GHz, 3.5\,GHz, and 10\,GHz, respectively. 
The goal of strategic RIS deployment is to improve RSRP at these outage locations, but deploying a dedicated RIS for each UE is infeasible due to the large number and distribution of outage UEs.

\begin{figure}[t]
    \centering
    \includegraphics[width=0.85\columnwidth]{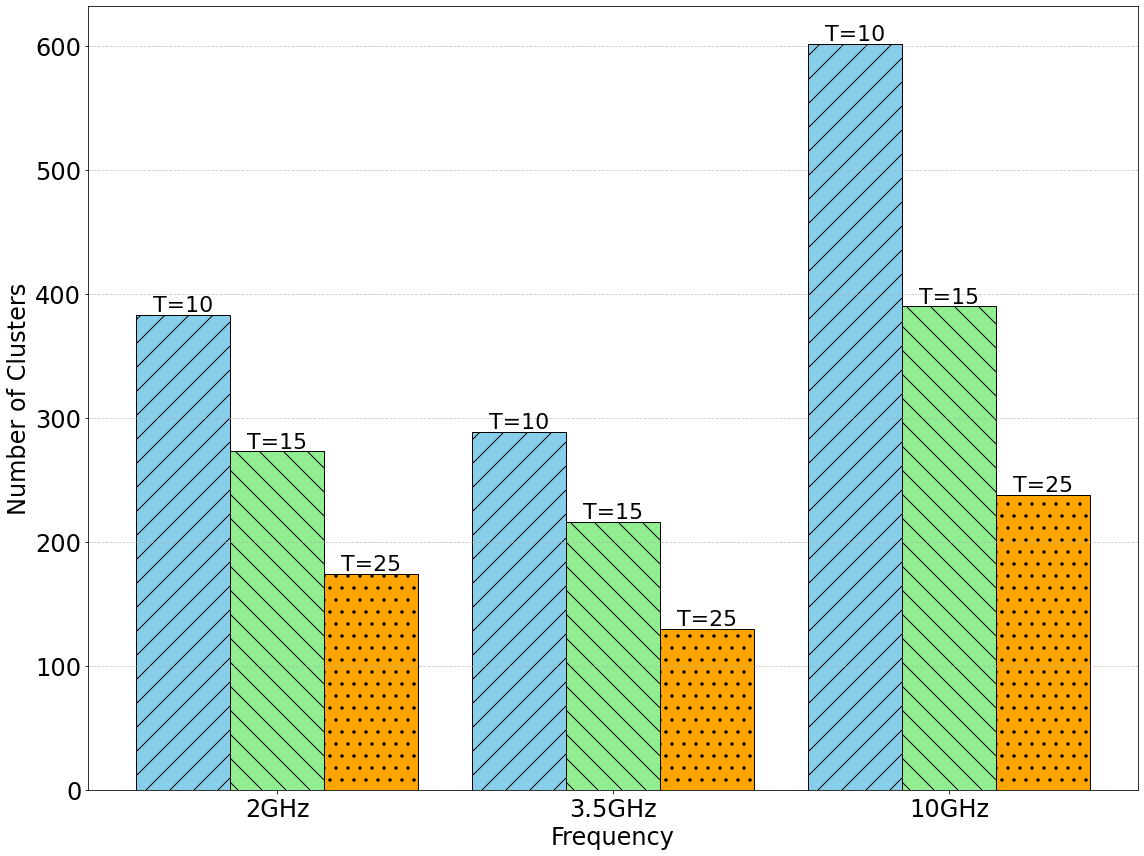}
    \caption{Number of clusters formed by the BIRCH algorithm vs. parameter \( T \) across the considered frequency bands.}
    \label{fig:number_clusters_vs_freq}
\end{figure}

To address this scalability challenge, nearby outage UEs are grouped into clusters so that each RIS serves a cluster instead of an individual UE. A hierarchical BIRCH algorithm~\cite{10.1145/233269.233324} is employed, building compact clustering feature trees. Each cluster is represented by
\begin{equation}
\left[ U, \sum_{i=1}^{U} \boldsymbol{s}_i, \sum_{i=1}^{U} \|\boldsymbol{s}_i\|^2 \right],
\label{eq:cluster_feature}
\end{equation}
where \( U \) is the number of UEs and \( \boldsymbol{s}_i \in \mathbb{R}^2 \) denotes the 2D position of UE \( i \) (center of a tile).
The threshold \(T\) is a key design parameter that defines the maximum allowable Euclidean distance (in meters) between data points within a cluster in the BIRCH algorithm. A smaller \( T \) yields more clusters and RISs, while a larger \( T \) reduces the count but enlarges coverage per RIS. Fig.~\ref{fig:number_clusters_vs_freq} illustrates how the cluster count changes with \( T \) across frequencies.

\subsection{RIS Candidate Locations}
\label{sec:reflection_based_approach}

This section presents the 
algorithm for the identification of candidate RIS locations.
For each cluster of outage UEs produced by BIRCH, one RIS is deployed to enhance coverage by aligning with the reflection point of relevant scattered rays. This study assumes RIS units can be deployed on any building surface with a valid scattering point, providing an upper bound on performance by ignoring practical constraints such as accessibility or regulations.

The 
approach identifies candidate RIS sites by activating only single-bounce scattering rays and increasing ray shootings to $3 \times 10^7$ for dense angular sampling. 

Among the 
rays that reach a given cluster centroid, the ones retained are those that (i) reach it after one scattering event and (ii) create a valid LoS path from the BS via the scatter point. Candidate RIS locations are then determined from the corresponding scatter points. If multiple candidates are found, the one with the shortest 3D distance to the cluster centroid is selected. If the centroid’s RSRP increases, UE-level gains within the cluster are assessed. A cluster is marked as RIS-effective if at least 60\% of its UEs experience an RSRP improvement. This balances RIS location efficiency with the computational complexity of the optimization. 

\subsection{RIS Configuration and BS Beamforming}
\label{sec:RIS_config_beamforming}

Once a candidate location is selected, the RIS must be oriented, configured, and incorporated into the network.

\begin{itemize}

\item Each RIS is installed along the building wall at the designated location, face outwards.

\item For each combination of BS, UE, and RIS locations, the spatial modulation function in Sec.~\ref{sec:RISmodel} is computed.

\item The serving BS selects the beam that maximizes the RSRP at the RIS location by iterating over all available DFT beams. 
\end{itemize}

\subsection{Re-Clustering and Re-Association}
\label{sec:reclustering}

\begin{figure*}[t]
    \centering
    \subfloat[4G at 2\,GHz]{%
        \includegraphics[width=0.32\textwidth]{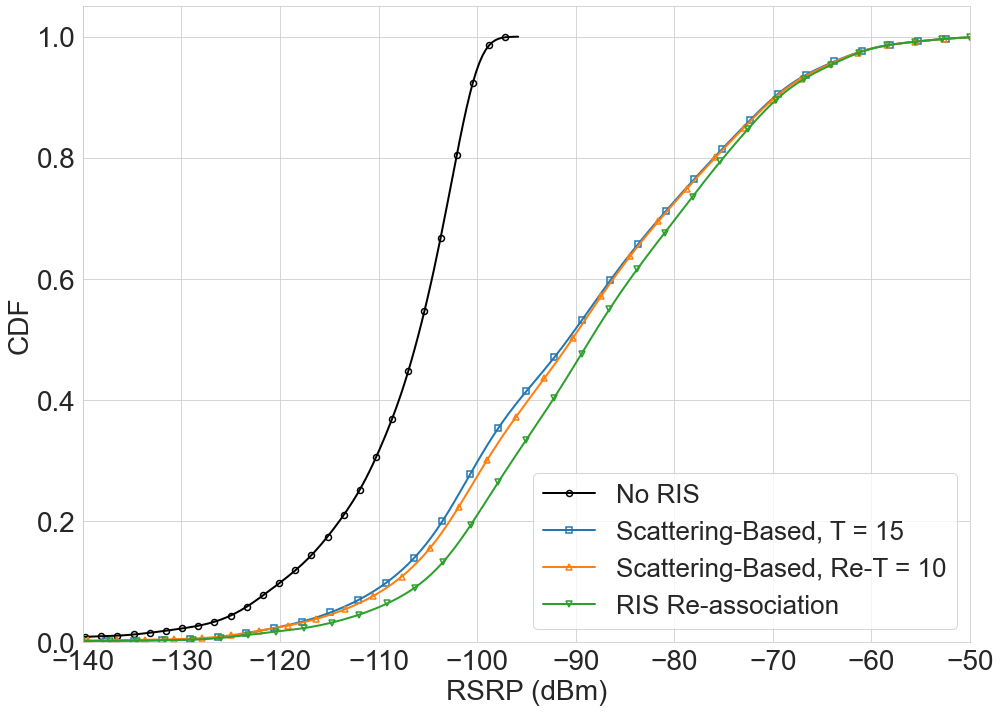}
        \label{fig:Optimization at 2GHz}
    }
    \hfill
    \subfloat[5G at 3.5\,GHz]{%
        \includegraphics[width=0.32\textwidth]{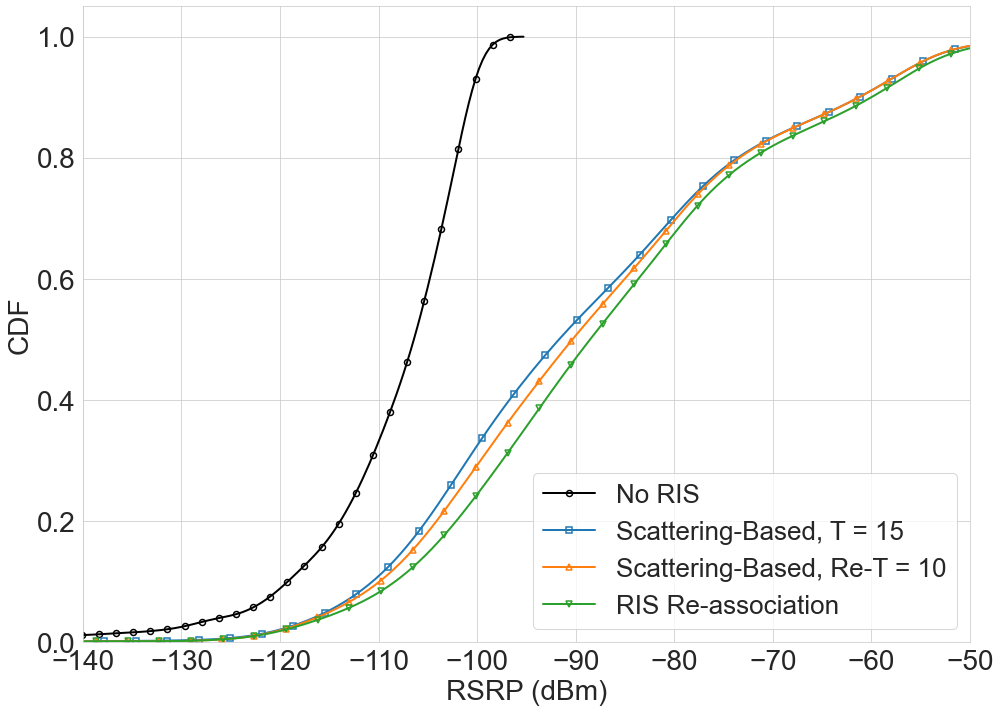}
        \label{fig:Optimization at 3.5GHz}
    }
    \hfill
    \subfloat[6G at 10\,GHz]{%
        \includegraphics[width=0.32\textwidth]{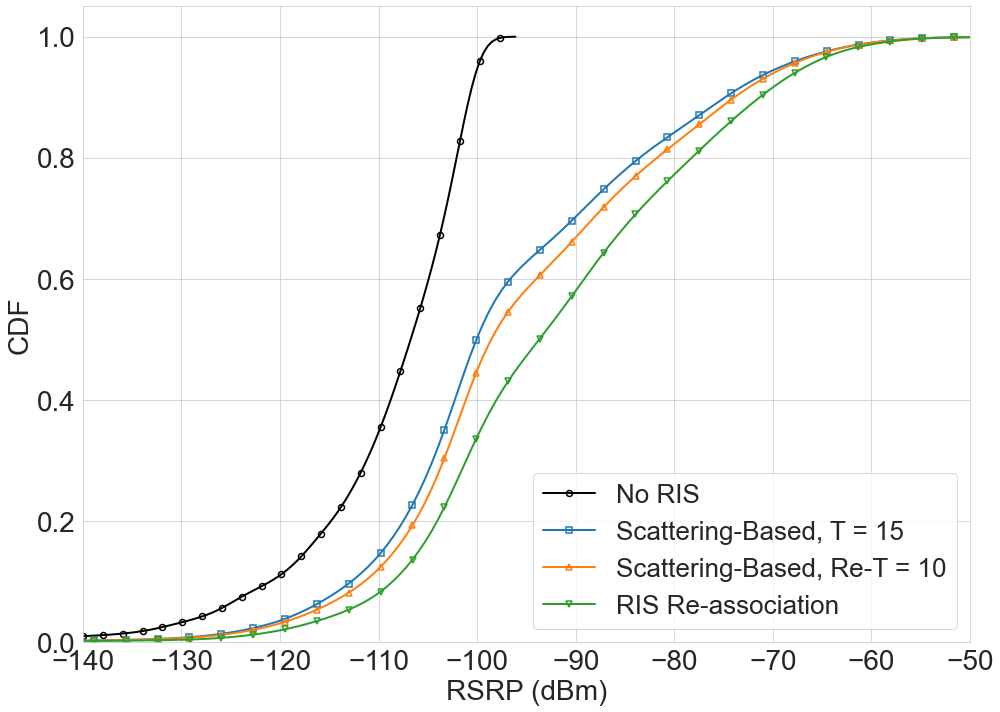}
        \label{fig:Optimization at 10GHz}
    }

    \vspace{0.5em} 

    \caption{RSRP enhancement from deploying the maximum number of RIS.}
    \label{fig:CDF_Comparison}
\end{figure*}

\begin{figure*}[t]
    \centering
    \subfloat[4G at 2\,GHz]{%
        \includegraphics[width=0.32\textwidth]{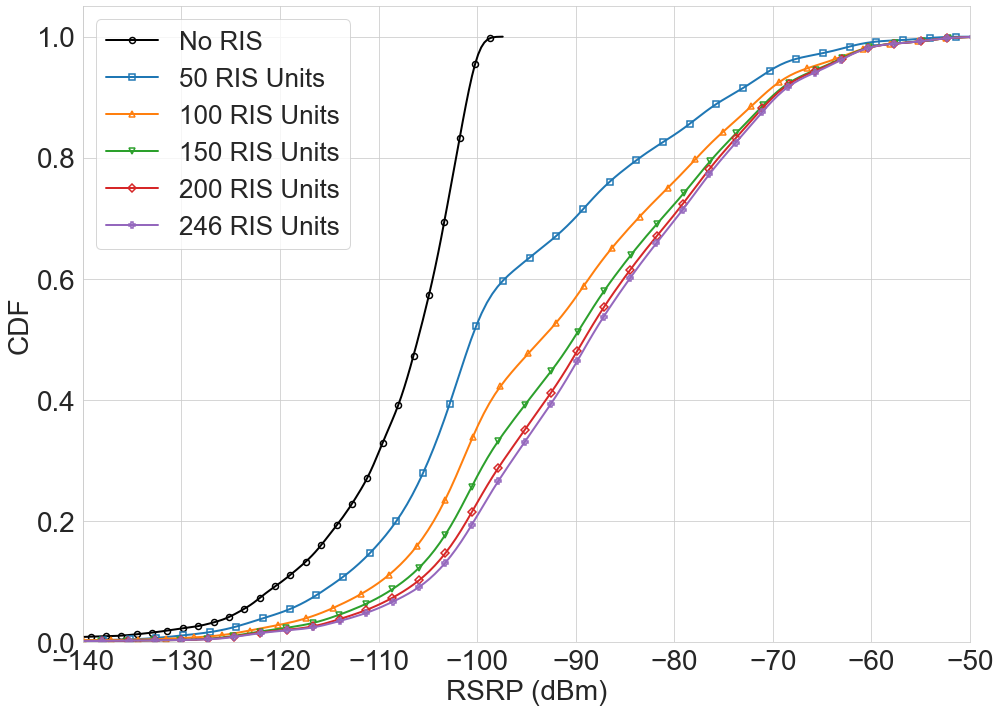}
        \label{fig:inc_RIS_2}
    }
    \hfill
    \subfloat[5G at 3.5\,GHz]{%
        \includegraphics[width=0.32\textwidth]{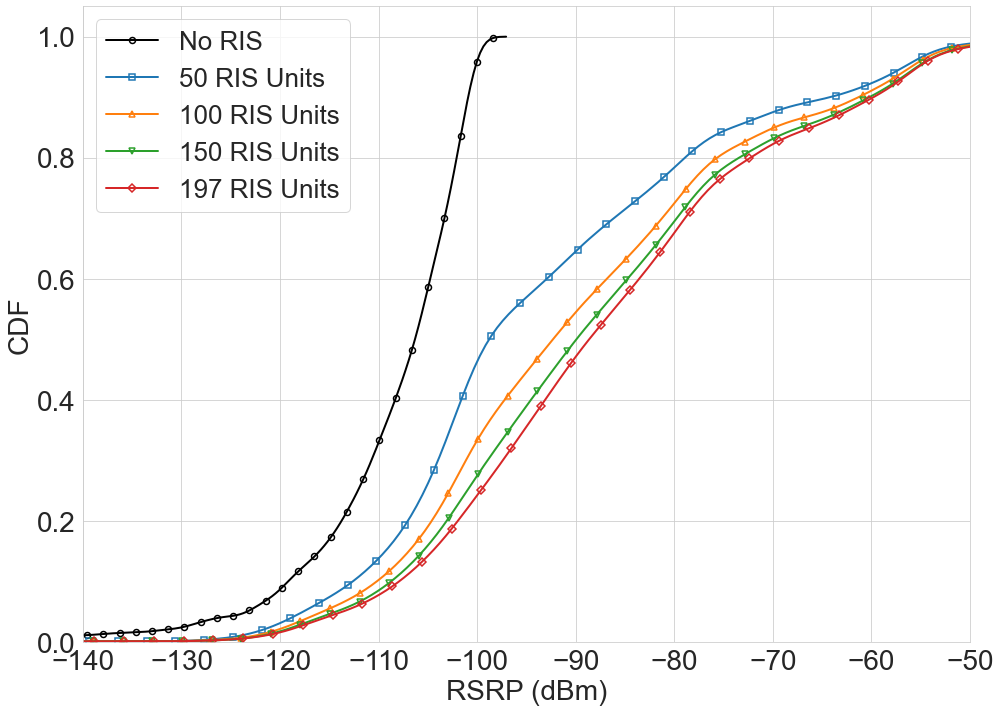}
        \label{fig:inc_RIS_3_5}
    }
    \hfill
    \subfloat[6G at 10\,GHz]{%
        \includegraphics[width=0.32\textwidth]{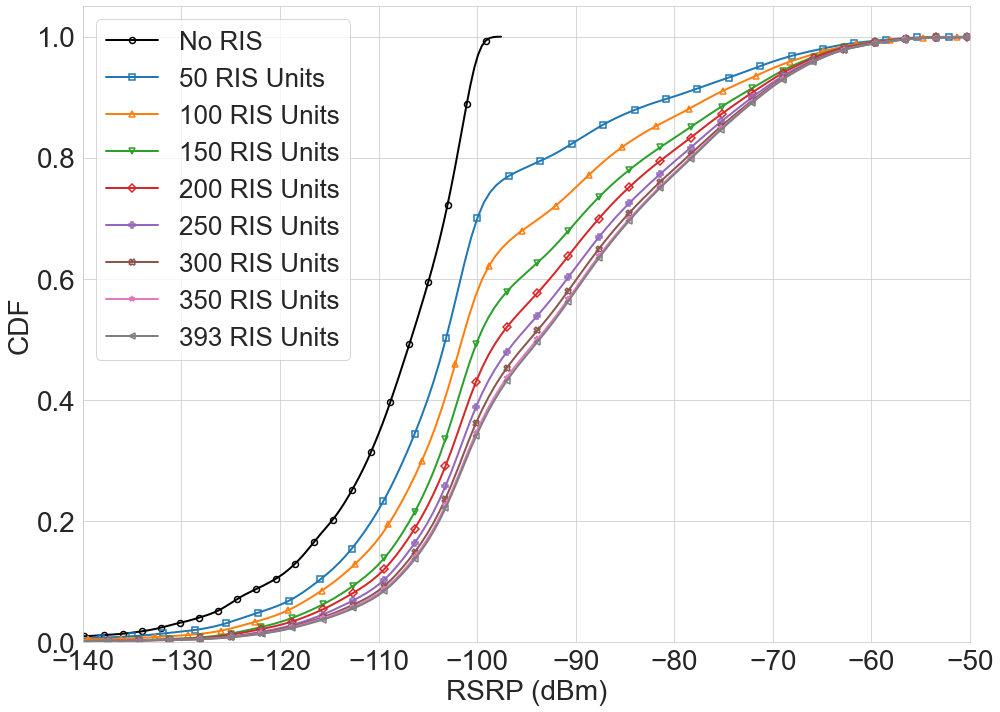}
        \label{fig:inc_RIS_10}
    }
    \caption{RSRP enhancement vs. number of deployed RIS.}
    \label{fig:Impact_RIS_incresing}
\end{figure*}

For the remaining outage UEs for which the candidate RIS locations proved ineffective, two fallback strategies are employed: re-clustering and RIS re-association.

\subsubsection{Re-Clustering}

The remaining outage UEs are regrouped with BIRCH using a smaller threshold \( T \) to form more compact clusters. The new threshold is determined iteratively to optimize the trade-off between coverage and deployment cost. For each new cluster, the centroid point is extracted, and the method is reapplied to find the RIS location. Although lowering \( T \) improves performance, it also increases the number of RIS units, raising capital costs for the MNO. 

\subsubsection{RIS Re-Association}

Some outage UEs may be in areas where existing RIS units are effective but were not initially assigned to their cluster. To avoid redundant deployments, these UEs can be reassigned to existing RISs if a LoS link is feasible.
The process consists of four steps:

\begin{enumerate}
    \item Determine which deployed RIS units maintain a LoS link to each outage UE.
    \item For each potential RIS, identify the BSs that have a LoS connection to it.
    \item From the feasible RIS options, select the one closest to the UE.
    \item From the BSs linked to the chosen RIS, pick the one nearest to it.
\end{enumerate}

This process refines the association of UEs, RISs, and BSs, boosting RSRP while avoiding further RIS deployments.

%% file: 05_Results.tex
\section{Simulation Results}
\label{sec:result}

\label{subsec:per_reflect}

For preliminary testing of the deployment method, a very large RIS aperture (11.24~$\times$~11.24 m) is considered, with half-wavelength element spacing. Outage UEs are clustered using BIRCH with a threshold of \( T = 15 \) m.

\subsubsection{4G at 2 GHz} Outage UEs represent 2.85\% of the total population. With \( T = 15 \)~m, 246 clusters are formed, each associated with one RIS. This initial RIS placement brings 68.85\% of outage UEs back into coverage. Re-clustering the residual UEs with \( T = 10 \)~m recovers an extra 2.73\%, while RIS re-association further improves coverage by 6.61\%, resulting in a total recovery of 78.97\% of outage UEs. Figure~\ref{fig:Optimization at 2GHz} illustrates the CDF of the enhanced RSRP values. 

\subsubsection{5G at 3.5 GHz} Outage UEs constitute 1.65\% of the entire network. Using the same method, RIS deployment initially restores coverage for 66.10\% of these UEs. Applying re-clustering with \( T = 10 \)~m adds another 3.80\%, while RIS re-association contributes an additional 5.67\%, resulting in an overall recovery of 75.25\%. Figure~\ref{fig:Optimization at 3.5GHz} shows these improvements, accomplished with 197 deployed RIS units. 

\subsubsection{6G at 10 GHz} At this frequency, 393 RIS units are deployed to address the 6.07\% of UEs initially in outage. The RIS deployment alone restores coverage for 46.02\% of these UEs, with re-clustering adding another 6.02\%. In particular, RIS re-association is especially effective in this scenario, recovering an additional 12.55\%, for a total of 64.59\% of outage UEs brought back into coverage, as shown in Fig.~\ref{fig:Optimization at 10GHz}.

While these initial findings show notable coverage gains, these entail a very large RIS aperture and many units, as each cluster is assigned a separate RIS. The remainder of this section evaluates how restricting aperture size and deploying RISs for only a subset of clusters influences performance.

\subsection{Impact of RIS Density}
\label{subsec:impact_RIS_density}


In the previous analysis, each cluster was associated with a separate RIS, resulting in 246, 197, and 393 RIS units at 2, 3.5, and 10\,GHz, respectively. However, many clusters are quite small---over half contain fewer than six UEs---reflecting the spatial sparsity of outage areas. To address this, clusters are now ranked by their UE count, and RIS units are incrementally allocated to the top-$N$ clusters as $N$ varies. Figures~\ref{fig:inc_RIS_2},~\ref{fig:inc_RIS_3_5}, and~\ref{fig:inc_RIS_10} show the resulting CDFs of RSRP for different values of $N$, showing that prioritizing larger clusters helps balance coverage gains against deployment costs. Nevertheless, even with prioritization, achieving significant improvements still needs deploying a substantial number of RIS units.

\subsection{Impact of RIS Apertures}
\label{subsec:impact_RIS_aperture}

\begin{figure}[t]
    \centering
    \includegraphics[width=\columnwidth]{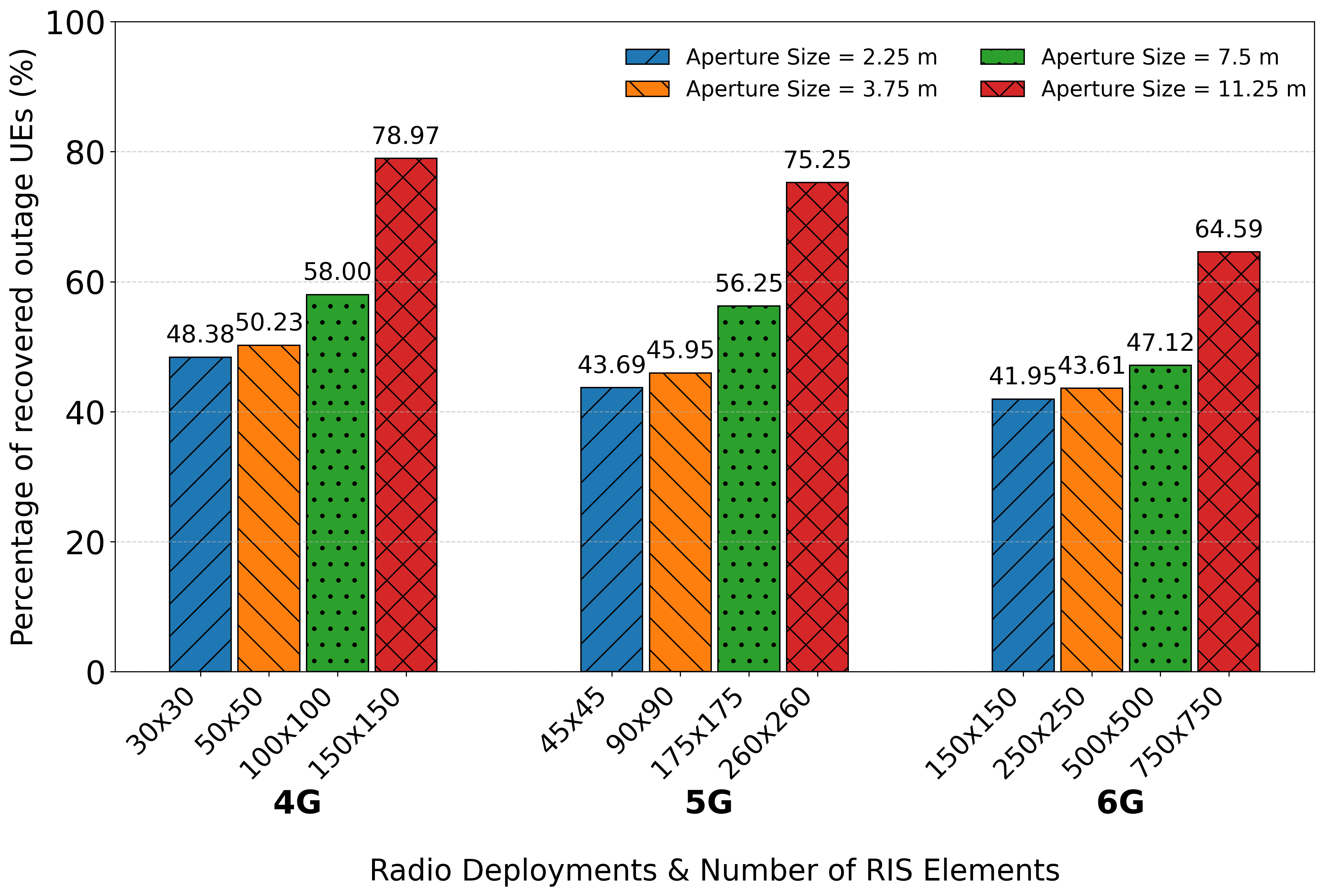}
    \caption{Percentage of outage UEs recovered with different RIS aperture sizes in various radio deployments.}
    \label{fig:ris_aperture}
\end{figure}

Next, we study the impact of varying the RIS aperture, which directly determines the number of RIS elements. The element spacing remains fixed at half a wavelength. This evaluation is conducted for the maximum number of RIS units previously identified, namely, 246, 197, and 393. Figure~\ref{fig:ris_aperture} presents the percentage of outage UEs recovered as a function of the aperture, with the corresponding number of RIS elements also indicated. The results show that performance is more robust to variations in aperture than to changes in RIS deployment density.

%% file: 06_Conclusion.tex
\section{Conclusion}
\label{sec:conclusion}

This work has introduced an automated, data-driven framework for evaluating large-scale RIS deployments in cellular networks. By integrating site-specific ray tracing, outage user clustering, and ray-based heuristics, the framework jointly determines RIS placement, orientation, configuration, and BS beamforming. Performance assessments in 4G, 5G, and 6G scenarios, grounded in a calibrated digital twin of an urban environment, highlight the tradeoff between coverage gains and deployment costs. Meaningful coverage enhancement necessitates deploying many large RIS units per square kilometer, raising questions about their cost-effectiveness in wide-area outdoor systems.